\documentclass[12pt]{iopart}
\usepackage{graphicx}
\begin{document}

\title[Holographic generation of complex fields]{Holographic
  generation of complex fields with spatial light modulators:
  Application to quantum key distribution}

\author{Mark T. Gruneisen$^1$, Warner A. Miller$^2$, 
  Raymond C. Dymale$^3$ and Ayman M. Sweiti$^2$}

     \address{$^1$ United States Air Force Research Laboratory,
       Directed Energy Directorate, Kirtland AFB, New Mexico 87117}
     \address{$^2$ Department of Physics, Florida Atlantic University, 
       Boca Raton, FL 33431}
     \address{$^3$ Boeing LTS, Inc., PO Box 5670, Albuquerque NM 87185} 

\date{16 May 2007}
 
\ead{Mark.Gruneisen@kirtland.af.mil, wam@fau.edu}
        
\begin{abstract} 
  There has been considerable interest recently in the generation of
  azimuthal phase functions associated with photon orbital angular
  momentum (OAM) for high-dimensional quantum key distribution (QKD).
  The generation of secure quantum keys requires not only this pure
  phase basis, but also additional bases comprised of orthonormal
  superposition states formed from the pure states.  These bases are
  also known as mutually unbiased bases (MUBs) and include quantum
  states whose wave functions are modulated in both phase and
  amplitude.  While modulo $2\pi$ optical path control with
  high-resolution spatial light modulators (SLMs) is well suited to
  creating the azimuthal phases associated with the pure states, it
  does not introduce the amplitude modulation associated with the MUB
  superposition states.  Using computer-generated holography (CGH)
  with the Leith-Upatnieks approach to hologram recording however,
  both phase and amplitude modulation can be achieved.  This paper
  presents a description of the OAM states of a 3-dimensional MUB
  system and analyzes the construction of these states via CGH with a
  phase modulating SLM.  The effects of phase holography artifacts on
  quantum-state generation are quantified and a prescription for
  avoiding these artifacts by preconditioning the hologram function is
  presented.  Practical effects associated with spatially isolating
  the first-order diffracted field are also quantified and a
  demonstration utilizing a liquid crystal SLM is presented.
\end{abstract}



\maketitle

\section{Introduction}
\label{sec:1}

With advancements in digital electronics, the principles of holography
\cite{1} continue to be utilized for new approaches to imaging and
diffractive control of optical fields.  In digital holography (DH), an
imaging sensor and computer record the hologram and image
reconstruction is accomplished via digital processing techniques to
present high resolution digital images.\cite{2} In computer-generated
holography (CGH), a hologram function can be generated computationally
and then displayed via digital printing, photolithography or
display-based technologies.\cite{1,3} A 2006 special issue of {\it
  Applied Optics} presents an overview of recent advances in DH and
CGH as applied to microscopy, optical trapping and tweezing,
interferometry, data recording, beam shaping, and metrology.\cite{4}

There has been considerable interest recently in the generation of
azimuthal phase functions associated with photon orbital angular
momentum (OAM) for high-dimensional quantum key distribution
(QKD).\cite{5} The complex fields associated with these functions are
of the form $\exp(il\theta)$ where $\theta$ is the azimuthal
coordinate.  The integer $l$ denotes the photon orbital angular
momentum in units of Planck’s constant over the range $-\infty < l
< \infty$.  While the conventional BB84 protocol for QKD \cite{6} is
based upon photon spin angular momentum and utilizes two polarization
bases (e.g. the horizontal and diagonal bases), extensions of this
protocol to higher dimensions is possible.\cite{7} Whether in two dimensions
or higher dimensions, secure transmission of quantum keys is ensured
by the use of mutually unbiased bases (MUBs).\cite{8,9} The states associated
with each MUB are equally weighted superpositions of the states
associated with the other bases. The measurement of a single photon’s
state with an incorrect choice of basis results in an unbiased, or
meaningless, probabilistic outcome.  Preserving this essential feature
of the BB84 protocol while implementing QKD with higher dimensional
OAM Hilbert spaces, requires one to generate and discriminate each
state in each of the MUBs associated with OAM.  These OAM MUB states
correspond to optical fields that are modulated in both phase and
amplitude.

Recent demonstrations5 of the generation of photon OAM have utilized
high-resolution programmable optical path modulators to impart the
azimuthal phase dependence of the pure states onto optical wavefronts.
For values of $l > 1$, the limited range of these spatial
light modulators (SLMs) requires that the optical phase function be
introduced modulo $2\pi$. Assuming monochromatic conditions and adhering
to the sign conventions of Goodman \cite{10} wherein a positive optical phase
delay introduces a negative phase error, the complex transmittance
associated with an optical phase delay function $l\theta$ introduced modulo
$2\pi$ can be shown to be \cite{11}-\cite{13} 
\begin{equation}
\label{1}
t(\theta) = \sum_{m=-\infty}^{\infty} sinc\left(1-m\right) e^{iml\theta}
\end{equation}
where $m$ is an integer and the sinc function is used according to the
convention $sinc(x)=\sin(\pi x)/\pi x$.  The amplitude diffraction
efficiency, given by the sinc function, is spatially uniform and unity
for $m=1$ and zero otherwise.  While modulo $2\pi$ optical phase control
is well suited to creating the azimuthal phases associated with the
pure states with high diffraction efficiency, the diffractive optics
function is calculated without the benefit of a reference wave to
encode amplitude information.

Methods for computing hologram functions that give rise to complex
field modulation include both analytic and numerical techniques.
Historically, the simplicity of fabricating binary holograms motivated
the development of numerical techniques for calculating and optimizing
the binary hologram function.\cite{1,14} Recent developments in SLM
technologies allow analytic hologram functions to be displayed in real
time with megapixel spatial resolution and $8$-bit phase resolution.
Consequently, analytic expressions for hologram recording functions
can be used directly in CGH.  The off-axis hologram recording approach
introduced by Emmett Leith and Juris Upatnieks \cite{15,16,17} defines
a useful approach to CGH with SLM technologies providing both phase
and amplitude modulation of the diffracted field and allowing
separation of the diffracted orders. This approach also allows the
hologram transmittance and the complex fields associated with the
diffracted orders to be expressed analytically for purpose of
analysis.

While photon polarization supports QKD in a $2$-dimensional Hilbert
space, in this paper we present a description of OAM MUBs for a
three-dimensional Hilbert space and consider the implications of
generating these quantum states via CGH with SLM technology. While our
analysis is valid for higher dimensional QKD, we present this
next-higher dimensional generalization for clarity.  Toward this end,
the complex transmittance associated with the Leith-Upatnieks approach
to recording thin phase holograms is reviewed and evaluated for the
case of holographic generation of a MUB state.  Artifacts in the
holographically generated complex field are discussed and their
effects on a QKD system quantified by calculating the weighted inner
product of the holographically generated field with the theoretical
MUB field.  Through preconditioning the CGH function, these artifacts
are eliminated.  Finally, a demonstration is presented in which an
extended graphics array (XGA) format liquid-crystal-on-silicon (LCOS)
SLM is utilized to holographically generate an optical field
associated with a three-dimensional OAM MUB state.

\section{Mutually Unbiased Bases of OAM States in 3 Dimensions}
\label{sec:2}

A 3-dimensional Hilbert space admits a maximum of four MUBs.  Each of
these bases is, by definition, orthonormal and contains three basis
vectors.  We may freely choose as one of these bases, designated
$MUB_0$, the three pure OAM states corresponding to an angular
momentum, $l_a=a\hbar$, $l_b=b\hbar$, and $l_c=c\hbar$, where $a,b,z
\in Z$.  To further reinforce the vector nature of these states, we
utilize Dirac’s ket notation \cite{18} to provide an abstract
description of each of these three basis vectors in this MUB.
\begin{equation}
\label{2}
MUB_0 = \left\{ | a \rangle, | b \rangle, | c \rangle \right\}
\end{equation}
These three ket vectors are orthogonal, providing $a\neq b\neq c$ ,
and span the $3$-dimensional Hilbert space.  Any other quantum state
$|\psi \rangle$ in this space can be written as a linear superposition
of the three basis vectors,
\begin{equation}
\label{3}
| \psi \rangle = \alpha | a \rangle + \beta | b \rangle + \gamma | c \rangle ,
\end{equation}
with $\alpha ,\_ \beta, \ \gamma \in C$ and
\begin{equation}
\label{4}
\alpha^* \alpha + \beta^* \beta + \gamma^* \gamma = 1
\end{equation}
We require that these states be normalized with inner product 
\begin{equation}
 \label{5}  
\langle i | j \rangle = 
\delta_{i,j}\ \ \hbox{for}\ i,\_j \in \left\{ a,b,c \right\}
\end{equation}
Consider a second MUB basis in this $3$-dimensional Hilbert space,
$MUB_1 = \left\{ |a_1 \rangle , | b_1 \rangle , | c_1 \rangle \right\}$.
Two characteristic features are needed to define another MUB basis.
First, the three vectors in such a basis must be orthonormal and
second, any state in $MUB_1$ should be equally distributed over the three
vectors in $MUB_0$.  Hence, 
\begin{equation}
\label{6}
| \langle i | j \rangle |^2 = \frac{1}{3}\ \ \hbox{for} i,j \in \{a_1,b_1,c_1\}
\end{equation}
The only four MUB bases in this $3$-dimensional Hilbert space are shown
in Table~\ref{table:1}.  Note that in Table~\ref{table:1},
$z=\exp(i2π/3)$ and we have suppressed the normalizing factor of
$1/\sqrt{3}$ in each of the nine MUB states in the last three columns.

\begin{table}
\begin{tabular} {| l |l | l | l | } \hline
$MUB_0$ & $MUB_1$ & $MUB_2$ & $MUB_3$ \\ \hline
$| a \rangle$ & 
$|a_1 \rangle \propto | a \rangle + | b \rangle + | c \rangle$ &
$|a_2 \rangle \propto | a \rangle + | b \rangle + z | c \rangle$ &
$|a_3 \rangle \propto | a \rangle + | b \rangle + z^2 | c \rangle$ \\
$| b \rangle$ & 
$|b_1 \rangle \propto | a \rangle + z | b \rangle + z^2 | c \rangle$ &
$|b_2 \rangle \propto | a \rangle + z | b \rangle + | c \rangle$ &
$|b_3 \rangle \propto | a \rangle + z | b \rangle + z | c \rangle$ \\
$| c \rangle$ & 
$|c_1 \rangle \propto | a \rangle + z^2 | b \rangle + z | c \rangle$ &
$|c_2 \rangle \propto | a \rangle + z^2 | b \rangle + z^2 | c \rangle$ &
$|c_3 \rangle \propto | a \rangle + z^2 | b \rangle + | c \rangle$ \\ \hline
\end{tabular}
  \caption{Dirac notation description of mutually unbiased bases in
    3-dimensional Hilbert space.}
\label{table:1}
\end{table}

Following the Dirac notation, we can provide a configuration-space
representation of each of these twelve azimuthal MUB states in terms
of a real-valued amplitude function $A(\theta)$ and a real-valued
phase function $\Phi(\theta)$.  The $MUB_0$ states are purely complex
and may be expressed as phase function in the optical plane.
\begin{eqnarray}
\label{7}
\langle \theta | a \rangle & = e^{ia\theta} \\
\langle \theta | b \rangle & = e^{ib\theta} \\
\langle \theta | c \rangle & = e^{ic\theta} 
\end{eqnarray}
Each of the other nine states in Table~\ref{table:1} is of the form, 
\begin{equation}
\label{8}
| d\rangle = \alpha | a \rangle + \beta | b \rangle + \gamma | c \rangle
\end{equation}
and accordingly, 
\begin{equation}
\label{9}
\langle \theta | d \rangle = A(\theta) e^{-i\Phi(\theta)}
\end{equation}
The corresponding real-valued amplitude and phase
functions can be expressed in terms of Eq.~\ref{9} as
\begin{eqnarray}
\label{10}
A(\theta)    & = \sqrt{ \langle d | d \rangle } \\
\Phi(\theta) & = -\tan{ \left( 
    \frac{
      Im\left(\langle \theta | d \rangle\right)}{
      Re\left(\langle \theta | d \rangle\right)}
    \right)}. 
\end{eqnarray}

For the purpose of this paper, and without loss of generalization, it
is convenient to assign the pure states $| a \rangle$, $| b \rangle$
and $| c \rangle$, the quantum numbers $| 1 \rangle$, $| 0 \rangle$
and $| -1 \rangle$, respectively.  Furthermore, the analysis that
follows will concentrate on CGH generation of one representative MUB
state, namely 
\begin{equation}
\label{11}
| c_3 \rangle = \frac{1}{\sqrt{3}} \left(
  | 1 \rangle + e^{-i 2\pi/3} | 0 \rangle + | -1 \rangle \right)
\end{equation} 
which may be expressed in complex field notation with the following
real-valued amplitude and phase, 
\begin{eqnarray}
\label{12}
A(\theta) & = \frac{1}{\sqrt{3}} \sqrt{1 - 2 \cos{\theta} + 4 \cos^2{\theta}} \\
\Phi(\theta) & = -\tan^{-1} \left( 
  \frac{\sqrt{3}}{4 \cos{(\theta)} - 1}
\right)
\end{eqnarray}

\section{Complex Field Modulation via Phase Modulation CGH}
\label{sec:3}

The off-axis hologram recording approach introduced by Leith and
Upatnieks defines a useful approach to complex field modulation by CGH
with SLM technology.  Introducing the reference wavefront off axis
leads to spatial separation of the diffracted orders allowing the
first-order diffracted field to be isolated from other diffracted
components.  The functional dependence of the complex field associated
with each diffracted order may be derived explicitly.  Fourier series
representations of thin phase hologram transmittance are given in the
standard references for the case of one-dimensional modulation
introduced by two planar wavefronts.\cite{1,10} In this section, we
present the more general form that includes spatial modulation of the
recorded field and show that phase modulation holography leads to
artifacts in both the amplitude and phase of the holographically
generated optical field.  Based on analytic expressions for these
artifacts, we present a prescription for preconditioning the
computer-generated hologram in order to avoid these artifacts in the
diffracted field.  The wavelength dependences associated with
holography are suppressed by assuming monochromatic conditions
throughout.

\subsection{Two-Dimensional Hologram Recording Function}

Let $a(\rho ,\theta)$ and $\phi (\rho ,\theta)$ represent spatially
varying real-valued amplitude and phase functions, respectively, where
$\rho$ and $\theta$ are the transverse radial and azimuthal coordinates,
respectively.  The complex optical field comprising both $a(\rho
,\theta)$ and $\phi (\rho ,\theta)$ may be written as
\begin{equation}
\label{13}
E(\rho,\theta) = a(\rho,\theta) e^{i\phi(\rho,\theta)}.
\end{equation}
Similarly, the phase associated with an off-axis reference wavefront
of slope $\alpha$ along the Cartesian $y$ direction is $k\,\alpha\,y$
which, in polar coordinates can be written as $kα\rho\sin{\theta}$, where
$k$ is the wave number.  Hence, the reference wave $R(\rho ,\theta)$ is
written as,
\begin{equation}
\label{14}
R(\rho,\theta) = b e^{-ik\alpha \rho \sin{\theta}},
\end{equation}
where $b$ is a constant amplitude.  In order to maximize fringe
contrast and maintain a monotonic relationship between fringe contrast
and amplitude, the reference beam amplitude $b$ is chosen to be equal
to the maximum value of the amplitude $a(\rho ,\theta)$ denoted
$a_{max}$.  The spatially varying intensity associated with the
superposition of both fields is
\begin{equation}
\label{15}
I(\rho,\theta) = a^2(\rho,\theta) + a^2_{max} + 2 a(\rho,\theta) a_{max}
\cos{(\phi(\rho,\theta)-k\alpha\rho\sin{\theta})}
\end{equation} .  
To allow for scaling of the hologram recording function, the
computer-generated holography function $f_{cgh}(\rho ,\theta)$ is defined
as
\begin{equation}
\label{16}
f_{cgh}\left(\rho,\theta\right) = \sigma \frac{I(\rho,\theta)}{I_{max}},
\end{equation}
where $I_{max}$ normalizes $I(\rho ,\theta)$ to unity and the
parameter $\sigma$ establishes the maximum value.

In principle, this function can be represented as a thin amplitude
hologram via an amplitude modulating SLM or as a thin phase hologram
via an optical path modulating SLM.  Commercially available optical
path modulating SLMs include piston-only micro-mirror arrays \cite{19}
based on micro electro-mechanical systems (MEMS) manufacturing and
parallel-aligned nematic liquid-crystal (PA-NLC) media when used with
linearly polarized light aligned with the NLC director.\cite{20}-\cite{24}
The variable birefringence of PA-NLC media also allows them to be used
in conjunction with orthogonal polarizers to create amplitude
modulation and is utilized in this manner for display applications.
It should be noted however that the complex transmittance also
includes a parasitic phase modulation term.  Marquez et al. \cite{25}
have shown that twisted NLC (TNLC) media can be configured with a
tandem arrangement of wave plates and polarizers to produce either
nearly pure amplitude modulation or nearly pure phase modulation.

\subsection{CGH with an Optical Path Modulating SLM }

For the case of implementation with an optical path modulating SLM,
the specific form of the complex transmittance is determined by the
phenomenology of the SLM.  In Sec.~\ref{5}, we will sample and scale the
CGH function of Eq.~\ref{16} in order to define an $8$-bit gray scale XGA
signal for the SLM driver.  The local response of the LC media is such
that the refractive index, and correspondingly the optical phase
retardance, decreases with increasing intensity as defined by Eq.~\ref{16}.
Therefore, the change in phase retardance introduced by the SLM is,
\begin{equation}
\label{17}
\Delta\phi_{SLM} (\rho,\theta) = -\sigma \frac{I(\rho,\theta)}{I_{max}}
\end{equation}
where $\sigma$ defines the maximum change in phase introduced by the
SLM.  The resulting complex transmittance $t(\rho ,\theta)$ of the hologram is
given by,
\begin{equation}
\label{18}
t(\rho,\theta) = e^{ -i \sigma \frac{I(\rho,\theta)}{I_{max}}}.
\end{equation}
By substituting Eq.~\ref{15} into Eq.~\ref{18} and applying the
Jacobi-Anger formula along with the identity\ref{26}
\[
J_m(-z)=(-1)^m J_m(z),
\]
the transmittance can be written in terms of the diffracted orders as,
\begin{equation}
\label{19}
t(\rho,\theta) =
e^{ -i \frac{\sigma'}{2}} 
\sum_{m=-\infty}^{\infty} 
(-1)^m 
J_m \left( \sigma' \frac{a(\rho,\theta) }{a_{max} } \right)
e^{ \left( i m (\phi(\rho,\theta)-k\alpha \rho \sin{\theta}) 
-i\frac{\sigma'}{2} \left( 
\frac{ a(\rho,\theta) }{ a_{max} } \right)^2 \right)}
\end{equation}
where $J_m$ is the Bessel function of the first kind of order m with
the maximum value of the argument given by $\sigma' = 2\sigma
a^2_{max}/I_{max}$.  The coefficient to the series is a spatially
uniform phase shift.  Assuming that the SLM is illuminated with the
reference wave of Eq.~\ref{14} with amplitude b now taken to be unity, the
transmitted optical field is given by the product of Eqs.~\ref{14} and 
\ref{19},
\begin{equation}
\label{20}
E_{out}\left(\rho,\theta\right) = e^{-i\frac{\sigma'}{2}} 
\sum_{m=-\infty}^{\infty} (-1)^m 
  J_m\left(\sigma'\frac{a(\rho,\theta)}{a_{max}}\right)
e^{-i \phi_m(\rho,\theta)},
\end{equation}
where the optical phase of the $m$-th order diffracted component is
\begin{equation}
\label{21}
\phi_m(\rho,\theta) = -m\phi(\rho,\theta) + (m+1) k\alpha\rho \sin{\theta} + 
\frac{\sigma'}{2} \left( \frac{a(\rho,\theta)}{a_{max}} \right)^2
\end{equation}

It is evident that the recorded phase $\phi (\rho ,\theta)$ is found
in the $m=-1$ order.  Applying the Bessel function identity
$J_m(z)=(-1)^mJ_m(z)$,\cite{26} the spatially varying amplitude of the
$m=-1$ diffracted order may be written as,
\begin{equation}
\label{22}
a_{-1}(\rho,\theta) = J_1\left(\sigma' \frac{a(\rho,\theta)}{a_{max}}  \right)
\end{equation}
The diffracted amplitude $a_1(\rho ,\theta)$ is related to the
recorded amplitude $a(\rho ,\theta)$ through the $J_1$ Bessel function
leading to nonlinearities in the amplitude of the holographically
generated optical field.  The spatially varying efficiency with which
optical power is diffracted into this order is given by the square of
$a_1(\rho ,\theta)$,
\begin{equation}
\label{23}
\eta_{-1}(\rho,\theta) = J_1^2\left(\sigma' \frac{a(\rho,\theta)}{a_{max}}  
\right),
\end{equation}
and achieves a maximum value of $33.9\%$ at the peak of the $J_1$
Bessel function.  If a given value for the amplitude of the diffracted
field is to be defined by a unique value of the argument of the Bessel
function, then the parameter $\sigma'$ should be chosen such that the
Bessel function remains monotonic, $\sigma' \leq 1.84$.  The upper
limit for the maximum intensity $I_{max}$ is $4a^2_{max}$.  For this
case, $\sigma' = \sigma/2$ and the Bessel function will remain
monotonic if $\sigma \leq 3.68$.

The spatially varying phase associated with the $m=-1$ diffracted
order, $\phi_{−1}(\rho ,\theta)$, is given by
\begin{equation}
\label{24}
\phi_{-1} (x,y) = \phi(\rho,\theta) + 
\frac{\sigma'}{2} \left( \frac{a(\rho,\theta)}{a_{max}} \right)^2
\end{equation}
While Eq.~\ref{24} indicates the desired phase $\phi (\rho ,\theta)$
is constructed by the hologram, it also shows the presence of the
aberration term $\frac{\sigma'}{2} \left(
  \frac{a(\rho,\theta)}{a_{max}} \right)^2$ .  This aberration term is
proportional to the spatially varying intensity $a^2(\rho ,\theta)$
and assumes its maximum value when $a(\rho ,\theta)=amax$.  For
$\sigma' = 1.84$, the maximum phase error would be $0.92$ radians or
$0.15$ waves, within the classical quarter-wave Rayleigh criteria for
diffraction limited performance.\cite{27}

For applications requiring a high degree of fidelity in the
holographically generated wavefronts, CGH offers the opportunity to
precondition the computed hologram to compensate for both the $J_1$
nonlinearity and the phase aberration.  In order to generate an
optical field with modulated amplitude $A(\rho ,\theta)$ it is only
necessary to define the CGH amplitude function $a(\rho ,\theta)$ such
that the amplitude of the $m=-1$ diffracted field satisfies the
relationship
\begin{equation}
\label{25}
a_{-1}(\rho,\theta) = J_1\left(\sigma' \frac{a(\rho,\theta)}{a_{max}}  \right)
= c A(\rho,\theta)
\end{equation}
where $c$ is a scaling factor that for a given value of $\sigma'$
matches the maximum values of $J_1$ and $A(\rho ,\theta)$.  Given the
preconditioned amplitude function $a(\rho ,\theta)$, it is then possible
to define the preconditioned phase function $\phi (\rho ,\theta)$ that
gives rise to $\Phi (\rho ,\theta)$.  This is done by setting 
$\phi_{−1}(\rho ,\theta)$ in Eq.~\ref{24} equal to $\Phi(\rho ,\theta)$.
\begin{equation}
\label{26}
\phi_{-1} (x,y) = \phi(\rho,\theta) + 
\frac{\sigma'}{2} \left( \frac{a(\rho,\theta)}{a_{max}} \right)^2 = 
\Phi(\rho,\theta)
\end{equation}

\section{Generation of an OAM MUB State via Phase Modulation
  Holography }
\label{sec:4}

The MUB states that we generate using CGH will never exactly represent
the theoretical MUB state.  In a QKD system, numerous effects will
introduce cumulative deviations from the theoretical state.  In
addition to the artifacts of phase modulation holography described
above, these can include optical aberrations within the system
including atmospheric aberrations.\cite{28} In this section, we
evaluate the effects of phase modulation holography artifacts on the
fidelity of a three dimensional MUB state.  We also consider the case
where the hologram is preconditioned to minimize these artifacts and
evaluate additional diffraction effects associated with spatially
isolating the first diffracted order.

\subsection{Artifacts of phase modulation holography}

\begin{figure}
\centering
\includegraphics[width=0.8\linewidth]{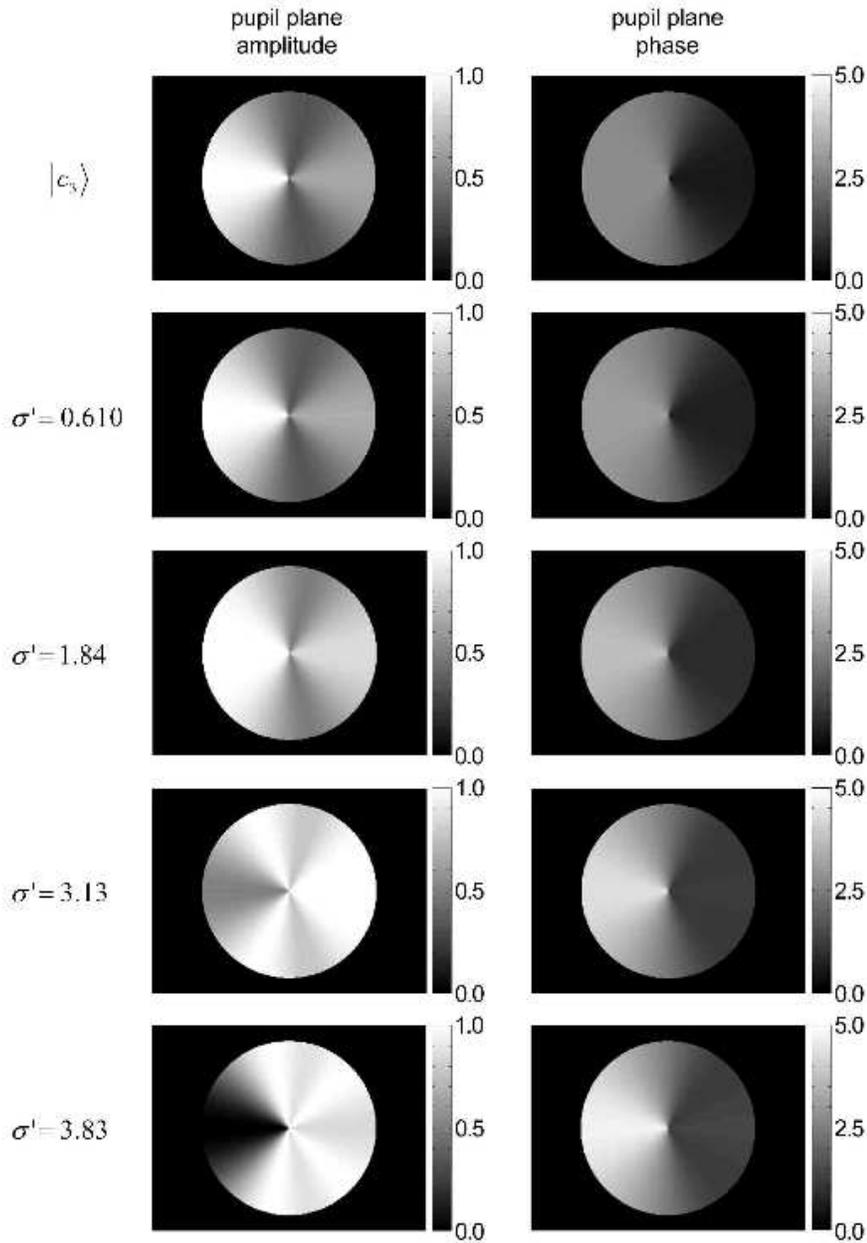}
\caption{Calculated amplitudes and phases comparing theoretical
  complex field to that generated by CGH with phase modulation
  holography for various values of the phase scaling parameter
  $\sigma'$. }
\label{fig:1}
\end{figure}

For the purpose of illustration, consider the $| c_3 \rangle$ state of
the previous section.  Recall from Eqs.~\ref{12} of Sec.~\ref{sec:2}
that the amplitude and phase associated with the superposition state
may be written as, 
\begin{eqnarray}
\label{27}
A(\theta) & = \frac{1}{\sqrt{3}} \left( 
1 - 2 \cos{(\theta)} + 4 \cos^2{(\theta)}
\right)^{\frac{1}{2}} \\
\Phi(\theta) & = -\tan{\left(\frac{\sqrt{3}}{4 \cos{(\theta)}-1}\right)}
\end{eqnarray}
These are shown in gray scale in the first row of Fig.~\ref{fig:1}
with the amplitude plot normalized to unity and the phase plot in
units of radians.  Taking the amplitude $a(\rho ,\theta)$ and phase
$\phi (\rho ,\theta )$ in the CGH calculation to be exactly those of
the $| c_3 \rangle$ state, $a(\rho ,\theta)=A(\rho ,\theta)$,
$amax=Amax$, and $\phi (\rho ,\theta )=\Phi (\rho ,\theta )$, we can
evaluate the effect of the phase modulation holography artifacts on
the holographically generated $| c_3 \rangle$ state.
Eqns.~\cite{22,24} give the amplitude and phase of the $m=-1$ order
diffracted field to be,
\begin{eqnarray}
\label{28}
a_{-1} &= J_1 \left( \sigma' \frac{A(\theta)}{A_{max}} \right) \\
\phi_{-1}(\theta) &= \Phi(\theta) + \frac{\sigma'}{2}\left(  
\frac{A(\theta)}{A_{max}}\right)^2
\end{eqnarray}
Upon substitution from Eqs.~\ref{28} the amplitude and phase of the
holographically generated field are,
\begin{eqnarray}
\label{29}
a_{-1} &= J_1\left(\frac{}{} \left( 
    1 - 2 \cos{(\theta)} + 4 \cos^2{(\theta)}
  \right)^{\frac{1}{2}}  \right) \\
\phi_{-1} &= -\tan{\left(\frac{\sqrt{3}}{4 \cos{(\theta)}-1}\right)} + 
\frac{\sigma'}{6 A^2_{max}} \left( 1 - 2 \cos{(\theta)} + 4 \cos^2{(\theta)}
  \right)
\end{eqnarray}
The amplitude and phase functions of Eqs.~\ref{29} are evaluated for
maximum values of the Bessel function argument given by $\sigma' =
0.610$, $1.84$, $3.13$, and $3.83$ corresponding to the first
half-maximum, peak, second half-maximum, and second zero of the Bessel
function, respectively.  The results are shown in gray scale in the
remaining rows of Fig.~\ref{fig:1}.  For the case of $\sigma' =
0.610$, the Bessel function is nearly linear and the first-order
amplitude is nearly identical to the $|c_3 \rangle$ state amplitude
shown in the first row.  For $\sigma' = 1.84$, the nonlinear scaling
is more significant and the effects become discernable in the
amplitude plot.  For larger values of $\sigma'$, the non-monotonic
nature of $a_1(\theta)$ is clearly evident.  At these larger values of
$\sigma'$, the phase aberration is also discernable in the plots.
Quantum mechanics provides us with a physically-meaningful metric over
the Hilbert space spanned by the apertured OAM quantum states $| l
\rangle$, where $l\in Z$ , that allows us to compare the fidelity, or
the angle between ket vectors, of our holographically generated
optical MUB states with the theoretical states.  In particular, the
metric takes the form of an inner product over the ket vectors between
our holographically produced MUB state $| \psi_h \rangle$ and the
theoretical MUB state $| \psi_t \rangle$, each of which can be
represented as wave functions in the azimuthal plane $\{\rho ,\theta\}$.
\begin{equation}
\label{30}
\langle \psi_t | \psi_h \rangle = 
\frac{
\int_{0}^{a} \int_{0}^{2\pi} 
\psi^*_t (\rho,\theta) \psi_h(\rho,\theta) \rho d\rho d\theta
}{| \psi_t |\, | \psi_h |}
\end{equation}
Here, $a$ is the aperture radius, and $| \psi |$ is the norm of the
wave function similarly defined as,
\begin{equation}
\label{31}
| \psi | = \sqrt{ \int_{0}^{a} \int_{0}^{2\pi} 
\psi^*(\rho,\theta)\psi(\rho,\theta) \rho d\rho d\theta }
\end{equation}
This inner product in the Hilbert space represents the cosine of the
angle in Hilbert space between our observed and idealized MUB state
but, more importantly, the square of this inner product represents the
probability $P$ that our holographically produced MUB state will be
measured (by a perfect MUB state detector) in the theoretical state.
\begin{equation}
\label{32}
P = | \langle \psi_t | \psi_h \rangle |^2
\end{equation}
We will examine the dependence of this fidelity measure on 1) the
phase parameter $\sigma'$ for the theoretically defined first
diffracted order and 2) the reference wavefront slope for numerical
simulations of generating and spatially isolating the first diffracted
order.

\begin{figure}
\centering
\includegraphics[width=0.8\linewidth]{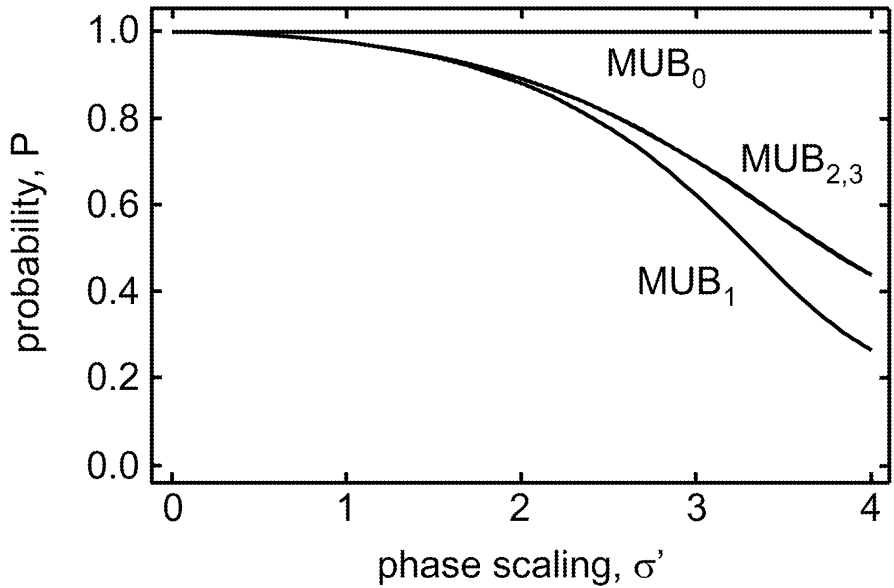}
\caption{Calculated inner-product-based probability quantifying the
  fidelity with which the complex field is generated via CGH vs. the
  phase scaling parameter $\sigma'$ for each of the mutually unbiased
  bases in 3-dimensional Hilbert space.}
\label{fig:2}
\end{figure}

Fig.~\ref{fig:2} shows the calculated probability function as a
function of the parameter $\sigma'$.  For sufficiently large values of
the tilt $\alpha$, the maximum intensity $I_{max}$ can be made
arbitrarily close to $4 A^2_{max}$.  For the purpose of
Fig.~\ref{fig:2}, it is sufficient to choose $\alpha$ to correspond to
$100$ waves of tilt across the aperture and let the maximum intensity
$I_{max}$ be represented by $4 a^2_{max}$.  The maximum value for the
argument of the Bessel function then becomes $\sigma' = \sigma/2$ and
the peak of $J_1$ occurs when $A(\theta)=A_{max}$ and $\sigma' = 1.84$
or, equivalently, $\sigma =3.68$.  The calculated probabilities for
all states within a given MUB are identical.  Since the pure states of
$MUB_0$ have constant amplitude, the phase aberration term of
Eqs.~\ref{28} results in a constant phase shift.  Similarly, the $J_1$
Bessel function is evaluated at a single value of the argument thus
avoiding nonlinearities in the amplitude response.  
Consequently, the
calculated probabilities for $MUB_0$ are unity for all values of
$\sigma'$.  The probabilities associated with the superposition states
of the remaining bases all decline with increasing $\sigma'$ as shown
with the states of $MUB_2$ and $MUB_3$ falling on the same curve.  At
$ \sigma' = 1.84$, the maximum value for which $J_1$ remains
monotonic, the probability functions have decreased to about $0.90$.
As $\sigma'$ increases through the range where $J_1$ is no longer
monotonic to $\sigma' =3.83$, the second zero of $J_1$, the
probability functions decrease further to the vicinity of $0.3$ to $0.5$
with the states in $MUB_1$ experiencing the largest reduction.

\subsection{Effects of Spatially Isolating the First-Order Diffracted Field}

Another practical effect associated with the holographic generation of
optical fields is that of isolating the first-order diffracted field
from other diffracted orders.  In practice, this is accomplished by
choosing the angle of the reference wavefront to be sufficiently large
that the angular diffraction is greater than the divergence associated
with the diffracted fields.  The first-order component may then be
transmitted by an aperture while blocking the other components.

\begin{figure}
\centering
\includegraphics[width=0.8\linewidth]{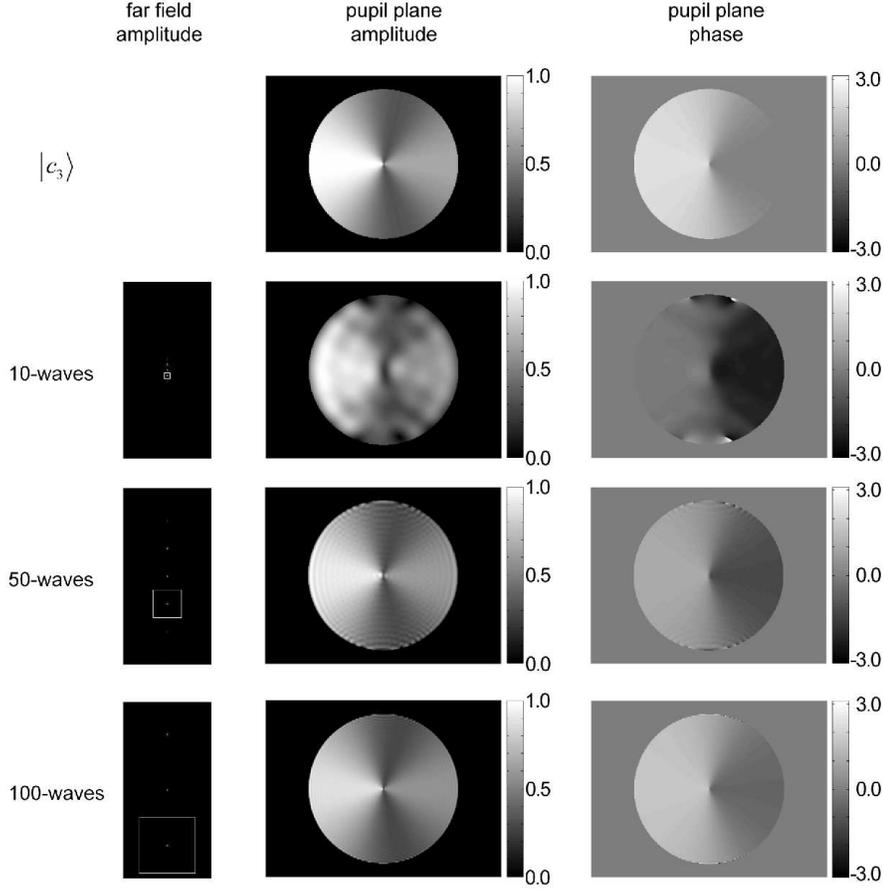}
\caption{ Numerically calculated amplitudes and phases resulting from
  holographic generation of complex field.  The calculation includes
  the effects of pre compensating phase holography artifacts and of
  spatially isolating the relevant diffracted order.  Several values
  of the reference wave tilt are shown.}
\label{fig:3}
\end{figure}

The effects of spatially isolating the holographically generated
$MUB_3$ field may also be evaluated numerically. The first row of
Fig.~\ref{fig:3} again shows the amplitude and phase associated with
the theoretical function as defined in Eqs.~\ref{27}.  Note the gray
scale for the phase plots is different than that in Fig.~\ref{1} in
order to remain consistent with the results that follow.  The second
row shows the numerically calculated holographically generated optical
field for the case of a reference wave with $10$-waves of tilt.  In
this case, the amplitude and phase artifacts of phase holography have
been avoided by using the preconditioned amplitude and phase functions
$a(\rho ,\theta)$ and $\phi (\rho ,\theta )$ as defined in
Eqs.~\ref{25},\ref{26} to calculate the CGH function of
Eq.~\ref{16}. The parameter $\sigma'$ is taken to be $1.72$ in order
that the Bessel function may be approximated by a polynomial and
Eqs.~\ref{25},\ref{26} may be solved for $a(\rho ,\theta)$ and $\phi
(\rho ,\theta )$, respectively.  The complex transmittance of
Eq.~\ref{18} is then used to define the optical field transmitted by
the hologram as elements of a $768\times 1024$ numerical array.  The
far-field amplitude is calculated by FFT and is shown in the first
column of Fig.~\ref{fig:3}.  The gray scaling has been chosen to be
nonlinear to emphasize the $m=\pm 1$ diffracted orders in the figure.
A rectangular aperture centered at the $m=-1$ diffracted order is
chosen such that the width of the aperture is equal to the order
spacing.  The field transmitted by the aperture is then transformed by
FFT back to the pupil plane.  The resulting amplitude and phase are
also shown in the second and third column, respectively.  Note that
while the orders appear to be well resolved, artifacts occur in the
amplitude and phase functions.  These artifacts may be attributed to
interference effects from adjacent orders and diffraction due to the
finite aperture size.  The third and fourth row show the effects of
increasing the reference wave tilt to $50$ and $100$ waves
respectively and scaling the aperture size accordingly.  Note that the
artifacts are significantly reduced.

\begin{figure}
\centering
\includegraphics[width=0.8\linewidth]{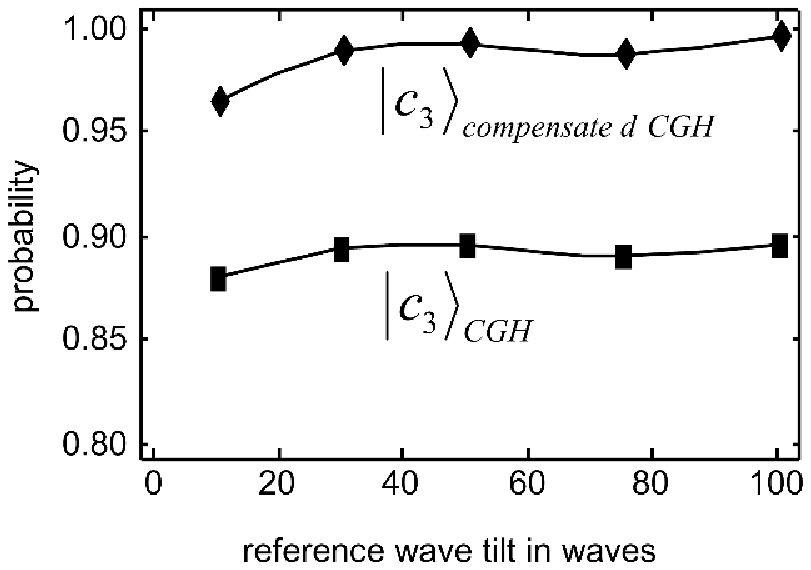}
\caption{ Calculated inner-product-based probability associated with
  the numerical calculation illustrated in Fig. 3.}
\label{fig:4}
\end{figure}

The fidelity of the holographically generated fields may again be
quantified by calculating the probability function in Eq.~\ref{32}
from numerical data similar to that shown in Fig.~\ref{fig:3}.  The
results are shown in Fig.~\ref{fig:4} for the case of the $| c_3
\rangle$ state and are calculated for the cases of $10$, $30$, $50$,
$75$, and $100$ waves of reference wave tilt.  The squares represent
the case where the CGH function was not preconditioned and therefore
include the effects of the amplitude and phase artifacts described in
Figs.~\ref{fig:1},\ref{fig:2}.  The diamonds represent the case in which the
artifacts are precompensated.  Fig.~\ref{fig:4} shows that the
combined effect of preconditioning the hologram and increasing the
reference tilt from $10$ to $100$ waves is to increase the
inner-product-based probability from $88\%$ to $99.6\%$.

\section{Experimental Demonstration with a LCOS SLM}

In this section, we utilize a LCOS SLM to demonstrate holographic
generation of the optical field associated with the $| c_3 \rangle$
quantum state described in the previous sections.  The LCOS SLM was
developed by Kent State Liquid Crystal Institute and Hana Microdisplay
under the Defense Advanced Research Projects Agency (DARPA) Terahertz
Operational Reachback (THOR) program.20 The LCOS SLM operates on the
principle of discrete-element optical path modulation over a $768\times
1024$ element XGA format.  The optical path length of each of the
elements may be independently varied over a range of more than $1$
micron corresponding to about two waves of modulation at the $532\,nm$
wavelength used in this demonstration.  Prior to using the LCOS SLM in
this demonstration, the nonlinearities in the optical phase response
were characterized and then utilized in a computer algorithm to
linearize the phase response to the CGH function.  In addition,
interferometric measurements of the SLM surface flatness were utilized
to generate a second aberration correction term in the preconditioned
CGH phase function $\phi (\rho ,\theta)$.  The preconditioned CGH phase
function is written as,
\begin{equation}
\label{33}
\phi(\rho,\theta) = \Phi(\theta) - 
\frac{\sigma'}{2} \left( k W(\rho,\theta) \right)
\end{equation} 
where $W(\rho ,\theta)$ is the wavefront error associated with the SLM
surface flatness and the parameter $\sigma'$ is again taken to be $1.72$.
The peak-to-valley magnitude of the SLM wavefront error is $2.15$
waves at $532\,nm$ optical wavelength. With $8$-bit resolution, the
phase response of the SLM spans a range of about $4\pi$ saturating at
a gray scale value of $190$.  The approximately $3.44$ radian phase
range used in this demonstration is accomplished over about $18$ phase
levels.  Details of these procedures have been published
previously.\cite{20},\cite{21}

\begin{figure}
\centering
\includegraphics[width=0.8\linewidth]{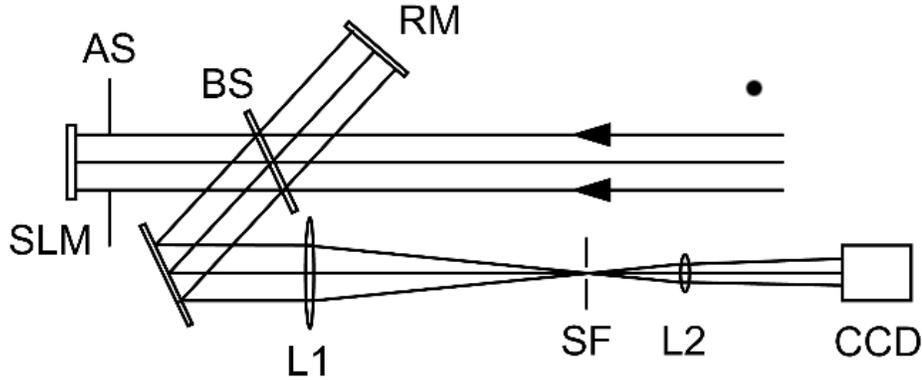}
\caption{ Schematic of experimental setup showing locations of
spatial light modulator (SLM), aperture stop (AS), beam splitter (BS),
reference mirror (RM), objective lens (L1), spatial filter (SF), field
lens (L2), and CCD camera.}
\label{fig:5}
\end{figure}

The experimental setup is shown in Fig.~\ref{fig:5}.  The output of a
$532\, nm$ continuous-wave frequency-doubled Nd:YAG laser is spatially
filtered, collimated to approximately $50\, mm$ beam diameter and
propagated to the LCOS SLM.  The aperture stop defines the area of the
SLM utilized to be $1\, cm$ or approximately $90\%$ of the active
area. The reflected/diffracted optical field is directed to a
one-meter focal length lens to create a far-field plane where another
aperture is inserted to spatially select the $m=-1$ order of the
hologram.  The transmitted order is then propagated to a $250 mm$
collimating lens that subsequently creates a re-imaged pupil plane
where a camera measures the pupil plane intensity distribution.  The
optical phase is measured interferometrically by inserting the
reference mirror as shown to create a planar reference wavefront and
form an interferogram on the camera.

\begin{figure}
\centering
\includegraphics[width=0.8\linewidth]{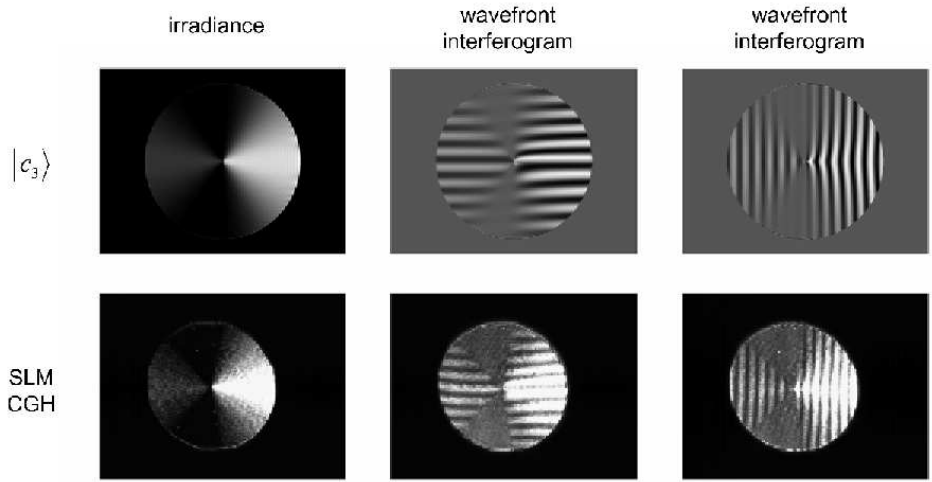}
\caption{Demonstration of state generation utilizing CGH and LCOS
SLM technology comparing calculated irradiance and interferograms to
those generated experimentally.}
\label{fig:6}
\end{figure}

The CGH function of Eq.~\ref{16} is calculated using $100$ waves of
reference wave tilt and using the phase function $\phi (\rho ,\theta)$
in Eq.~\ref{33} and the amplitude function $a(\rho ,\theta)$
satisfying Eq.~\ref{25} with $A(\rho ,\theta)$ and $\Phi (\rho
,\theta)$ given by the $| c_3 \rangle$ optical field as given in
Eqs.~\ref{27}.  The CGH function is scaled to produce a phase range of
$3.44$ radians on the SLM and sampled to define a $768\times 1024$ XGA
signal that is sent to the SLM driver.  The experimentally generated
optical field may be compared to the theoretical $| c_3 \rangle$ state
as follows.  The theoretical irradiance function is calculated form
the square of the $| c_3 \rangle$ state amplitude in Eqs.~\ref{27}.
Interferograms associated with the theoretical $| c_3 \rangle$ optical
field are calculated by summing the complex field of Eqs.~\ref{27}
with a planar reference wavefront and finding the squared modulus.
This is done for $10$ waves of interferometer reference tilt in both
the vertical and horizontal dimensions as shown in the first row of
Fig.~\ref{fig:6}.  The second row of Fig.~\ref{fig:6} shows the
experimentally measured irradiance and interferograms corresponding to
the theoretical case in the first row.  The experimental results are
in qualitative agreement with the analysis presented in this paper.

\section{Conclusions}

Computer-generated holography with spatial light modulator technology
is evaluated for its utility in generating QKD states based on photon
orbital angular momentum.  Utilizing azimuthal phase functions to
define the orthonormal pure states of a $3$-dimensional basis, the
mutually unbiased bases of a $4$-bases system are calculated and shown
to be azimuthal amplitude and phase functions of the optical field.
An analysis of CGH with $2$-dimensional phase modulators shows that
these optical fields may be holographically generated and that
artifacts associated with thin phase holography may be compensated in
the computation of the hologram.  Furthermore, a numerical analysis of
practical effects associated with spatially isolating the
holographically generated field defines a regime in which the
azimuthal field may be constructed with high fidelity.  A
demonstration with a LCOS SLM shows practical implementation of this
approach to generating quantum states for QKD with commercially
available technology.  Some effects associated with SLM technology,
such as pixilation and discretization of phase levels, were not
included in the analysis presented here.  These details are important
in the optimization of SLM designs for implementation of OAM-based
QKD.

\ack The authors gratefully acknowledge the important contribution of
Prof. Phil Bos, Kent State Liquid Crystal Institute, for his role in
both developing and allowing us the use of the LCOS SLM unit described
in this work.  We also wish to acknowledge important discussions with
Glenn Tyler of The Optical Sciences Company and Chris Beetle of
Florida Atlantic University.  We wish to thank Robert Fugate of the
Air Force Research Laboratory and L. N. Durvasula of DARPA for
suggesting this research topic.  This work was supported by the Air
Force Office of Scientific Research.

\end{document}